\documentclass{emulateapj}

\usepackage{longtable}
\usepackage{hyperref}
\usepackage{afterpage}
\usepackage{footmisc}

\shorttitle{Multiple Stellar Populations on Globular Clusters}
\shortauthors{Soto et al.}

\begin{document}


\title{The Hubble Space Telescope UV Legacy Survey of Galactic
  Globular Clusters. \\ 
VIII,  Preliminary Public Catalog Release}


\author{M.  Soto\altaffilmark{1,2}} 
\author{A. Bellini\altaffilmark{1}}
\author{J. Anderson\altaffilmark{1}}
\author{G. Piotto\altaffilmark{3, 4}}
\author{L. R. Bedin\altaffilmark{4}}
\author{R. P. van der Marel\altaffilmark{1}}
\author{A. P. Milone\altaffilmark{5}}            
\author{T. M. Brown\altaffilmark{1}}
\author{A. M. Cool\altaffilmark{6}}
\author{I. R. King\altaffilmark{7}}
\author{A. Sarajedini\altaffilmark{8}}
\author{V. Granata\altaffilmark{3,4}}
\author{S. Cassisi\altaffilmark{9}}
\author{A. Aparicio\altaffilmark{10,11}}
\author{S. Hidalgo\altaffilmark{10,11}}
\author{S. Ortolani\altaffilmark{3,4}}
\author{D. Nardiello\altaffilmark{3,4}}



\altaffiltext{1}{Space Telescope Science Institute, San Martin Drive 3700,
 Baltimore, 21218}\email{mario.soto@uda.cl}
\altaffiltext{2}{Universidad de Atacama, Departamento de F\'isica, Copayapu 485,
Copiap\'o, Chile}
\altaffiltext{3}{Dipartimento di Fisica e Astronomia “Galileo Galilei”, Universit\`a 
di Padova, Vicolo dell'Osservatorio 3, I-35122 Padova, Italy}
\altaffiltext{4}{INAF-Osservatorio Astronomico di Padova, Vicolo
  dell'Osservatorio 5, I-35122 Padova, Italy}
\altaffiltext{5}{Research School of Astronomy and Astrophysics, The 
Australian National University, Cotter Road, Weston, ACT, 2611, Australia}
\altaffiltext{6}{Department of Physics and Astronomy, San Francisco
  State University, 1600 Holloway Avenue, San Francisco, CA 94132}
\altaffiltext{7}{Department of Astronomy, University of Washington, Box 351580, Seattle, WA 98195-1580}
\altaffiltext{8}{Department of Astronomy, University of Florida, 211 Bryant Space Science Center, Gainesville, FL 32611, USA}
\altaffiltext{9}{Osservatorio Astronomico di Teramo, Via Mentore Maggini s.n.c., I-64100 Teramo, Italy}
\altaffiltext{10}{Instituto de Astrof\'isica de Canarias, E-38200 La
  Laguna, Tenerife, Canary Islands, Spain}
\altaffiltext{11}{Department of Astrophysics, University of La Laguna,
  E-38200 La Laguna, Tenerife, Canary Islands, Spain}
%


\begin{abstract}
The \emph{ Hubble Space Telescope UV Legacy Survey of Galactic
  Globular Clusters} (GO-13297) has been specifically designed to
complement the existing F606W and F814W observations of the 
ACS Globular Cluster Survey (GO-10775) by observing the most accessible 
47 of the previous survey's 65 clusters in three WFC3/UVIS filters 
F275W, F336W, and F438W.  The new survey also adds
 super-solar metallicity
open cluster
NGC~6791 to increase the metallicity diversity.
The combined survey provides a homogeneous 5-band data set that 
can be used to pursue a broad range of scientific investigations.  In
particular, the chosen UV filters allow the identification of multiple 
stellar populations by targeting the regions of the spectrum that are 
sensitive to abundance variations in C, N, and O.  In order to provide
the community with uniform preliminary catalogs, we have devised an 
automated procedure that performs high-quality photometry on the new UV
observations (along with similar observations 
 of  seven other programs
in the archive).  
This procedure finds and measures the potential sources on each individual
exposure using library point-spread functions and cross-correlates these
observations with the original ACS-Survey catalog.
The catalog of 57 clusters we publish here will be useful to identify
stars in the different stellar 
populations, in particular for spectroscopic follow-up.
Eventually, we will construct a more sophisticated catalog and artificial-star tests
based on an optimal reduction of the UV survey data, but the catalogs 
presented here give the community the chance to make early use of 
this {\it HST} Treasury survey.

%
%

\end{abstract}





%
%


\section{Introduction}

Early in the 20th century, Harlow Shapley used the distribution 
of Globular Clusters (GCs) to determine the extent of the Milky Way 
(Shapley 1918) and the Sun's position in it to overturn the most
accepted model of the Galaxy at the time, the Kapteyn Universe.  
Nowadays, GCs are still recognized as an important 
component of
Galactic structure and hold important clues for numerous questions related 
to stellar evolution, stellar dynamics, variable and binary stars, 
and the conditions present during the formation of the Milky Way itself.  
%
Consequently, over the years, important efforts have been made 
to construct complete GC catalogs and thus address some of these questions 
(e.g. Shapley et al.\ 1930; Arp 1965; Webbink 1985; Djorgovski \& King 1986; 
Djorgovski \& Meylan 1993; Harris 1996).  However, many of the properties 
included in these catalogs were obtained from integrated light or small 
stellar samples, 
and that 
necessarily constrained the scope of resulting science
to general parameters such as total brightness, colors, metallicities and 
reddenings.  In addition, many of the early catalogs were limited by 
the inhomogeneity that results from different instruments and set-ups.

GC surveys based on homogeneous data-sets are highly desirable since 
that facilitates comparisons such as those involving age or
horizontal-branch morphology. 
Initial attempts to provide homogenous 
samples, such as those of Rosenberg et al.\ (2000), which produced a catalog 
of 56 clusters using ground-base data,  and Piotto et al.\ (2002), which
sampled the cores of 74 clusters using Wide Field Planetary Camera 2 (WFPC2), 
hinted at what such homogeneous data-sets could do.  It was only natural  
that {\it HST}'s Advanced Camera for Surveys (ACS) would contribute the most uniform,
homogeneous catalog to date in the ACS Globular Cluster Survey 
(GO-10775; Sarajedini et al.\ 2007 and Anderson et al.\ 2008).  This survey
improved considerably on many of the limitations of previous surveys, 
which typically suffered from crowding in the central regions or
uneven 
sampling (e.g. the differences between the PC and the WF chips 
in the WFPC2).  The legacy value of the ACS Globular Cluster Survey  
(henceforth ACS GCS) has been 
%
demonstrated by studies addressing a broad range of topics related to GCs, 
such as mass functions (Paust et al.\ 2010), horizontal-branch morphologies 
(Dotter et al.\ 2010; Milone et al.\ 2014), relative ages
(Mar\'in-Franch et al.\ 2009; VandenBerg et al.\ 2013), the 
Sagittarius dwarf spheroidal galaxy nucleus (Siegel et al.\ 2007), stellar evolution 
tracks and isochrones (Dotter et al.\ 2007), 
 relative ages of GCs (De Angeli et al. 2005),   
clusters with multiple 
sequences (e.g. NGC~1851; Milone et al.\ 2008,  Piotto et al.\ 2012, Piotto et al.\ 2015), 
and mass segregation (Goldsbury et al.\ 2013).

Even at the time of the ACS GCS, GCs were largely assumed to be the
quintessential examples of 
simple
stellar populations, born in a single
starburst episode.  Anderson (1997), Lee et al. (1999), and Bedin et al.\ (2004) provided 
the first counter-example to this rule by showing that the main-sequence
of $\omega$ Centauri was divided into at least two distinct sequences.  Even
though hints of this had been seen spectroscopically and 
photometrically along the RGB, 
the main-sequence data showed that this was not simply an artifact of 
atmospheric evolutionary mixing.  
%

This initial discovery was soon followed 
by evidence of multiple 
populations (MPs) in other GCs: 
NGC~2808 (D'Antona et al.\ 2005; Piotto et al.\ 2007), 
NGC~6441 (Gratton et al.\ 2006; Bellini et al.\ 2013),
NGC~1851 (Milone et al.\ 2008; Cassisi et al.\ 2008), 
NGC~6388 (Carretta et al.\  2007; Bellini et al.\ 2013, Piotto et al.\ 2015); 
NGC~6121 (M4; Marino et al.\ 2008), 
NGC~6656 (M22; Marino et al.\ 2009), 
NGC~104  (47 Tuc; Anderson et al.\ 2009; Milone et al.\ 2012) 
NGC~6752 (Milone et al.\ 2010; Milone et  al. 2013), 
NGC~2419 (Di Criscienzo et al.\ 2011), 
and NGC~288 (Piotto et al.\ 2013). 
In many cases the detection of the MPs was made through the Na-O 
anti-correlation that stems from the CNO-cycle processed differences 
of second generation (2G) stars compared with first generation (1G) stars 
(Milone et al.\ 2013).  
The  
%
filters F275W, F336W, and 
F438W 
of the Wide Field Camera 3 (WFC3) in its Ultraviolet-Visible (UVIS) channel  
happen to be particularly sensitive to these differences.
While F275W covers the OH molecular band, F336W includes an NH band, 
and F438W CN and CH bands.  Hence, 1G stars are C- and O-rich and N-poor 
 and tend to be faint in F275W and F438W, while remaining bright in F336W.  
On the contrary, 2G stars are C- and O-poor and N-rich and tend to be  bright 
in F275W and F438W, but faint in F336W  (Milone et al.\ 2012). 
Since this filter set is so perfectly suited to study multiple populations 
(as well as many other cluster phenomena), the \emph{Hubble Space Telescope UV Legacy 
Survey of Galactic Globular Clusters} (GO-13297, PI-Piotto),  
was designed to use these filters in order  
to complement the existing optical-based ACS GCS.

Piotto et al.\ (2015) describe the motivation and observing strategy for 
the survey and show some preliminary results in terms of color-magnitude diagrams.  
The current paper will provide a preliminary catalog of the survey results.  
It is organized as follows:  in Section 2 we describe the WFC3/UVIS 
observations that have been used to construct each cluster catalog.  
In Section 3, we give a detailed description of our procedure, including 
the detection and measuring processes as well as the 
cross-identification
with the ACS GCS catalogs.  We discuss the advantages 
and shortcomings of these preliminary catalogs in Section 4.  Finally, 
Section 5 summarizes the current catalog and points to an upcoming final 
version.

%
%

\section{Observations}





As mentioned above, the GO-13297 UV survey has been designed primarily as a 
follow-on to the existing $F606W$ and $F814W$ observations of the ACS 
GCS (see Sarajedini et al.\ 2007).  The ACS GCS
sample of clusters was originally selected based on several criteria, the 
most important being proximity to the Sun and low reddening [$E(B-V)< 0.35$].
With one exception, we restricted our sample to the clusters observed 
in the ACS GCS, but had to exclude those that were either too distant 
or too reddened to make observations in the UV practical.  We also excluded 
 three of the clusters that had already been observed in the UVIS filters.
In all, we included 47 of the 65 ACS GCS clusters and added NGC~6791, 
an old metal-rich open cluster  
%
 that extends our multiple stellar population investigation
 to super-solar metallicity.

As described more fully in Piotto et al.\ (2015), the faintness of stars in
the F275W band is the main limitation to the survey.  It would be prohibitive to
get S/N $\sim$ 10 for all stars in Sarajedini's survey in the UV bands, so we aimed 
simply to reach a minimum accuracy of $0.02$ magnitude (or $S/N \gtrsim 50$ ), 
in the F275W band down to the turnoff.  A total of 131 orbits were allocated 
to this program:  15 clusters needed only 2 orbits to attain four exposures per 
band according to our specifications, 4 clusters required 3 orbits each, 
8 clusters were observed in 4 orbits, 3 clusters had 5 orbits, 2 clusters 
required 6 orbits, while 2 others only had 1 orbit each in only two bands.  
The remaining 14 clusters are either too distant or too reddened to detect 
MPs for stars below the turn-off, so for them we assigned just 2 orbits 
 to allow identification of the MPs in the evolved populations. 

Previous programs with suitable WFC3/UVIS observations in F275W, F336W,
and F438W have also contributed to the catalog we present here.  Pilot  
 MPs WFC3/UVIS programs GO-12311 (PI: G. Piotto), GO-12605 
(PI: G. Piotto),   
and the programs with similar observations, GO-11633 (PI: R.M. Rich), 
GO-11729 (PI: J. Holtzman), GO-12008 (PI: A. Kong), GO-12746 (PI: A. Kong),
 and GO-12971 (PI: H. Richer), 
allowed the sample presented here to be expanded 
to 57 clusters with 5-band data. 

In addition to these WFC3/UVIS observations, described above, parallel 
ACS field observations have also been carried out for each cluster in the 
F475W and F814W bands.  These parallel observations 
were designed 
to overlap with archival data in the outer regions of each cluster as much 
as possible, thus providing the foundation for proper-motion studies at 
different radii for a significant number of these clusters.  The catalog 
presented here will focus exclusively on the core region observed by WFC3/UVIS.

A detailed account of the exposures for each 
cluster
can be found in 
Piotto et al.\ (2015).  For convenience, we provide a brief summary in
Table~\ref{table:tab1}.

\begin{table*}
\begin{center}
\caption{Summary of Cluster Observations \label{table:tab1}}
\begin{tabular}{crrlllc}
\tableline\tableline
Cluster & RA \ \ \  & DEC \ \ \   & F275W & F336W & F438W  & $\Delta$Epoch (yrs) \\
\tableline

NGC 0104 & 00:24:15 & $-$72:05:47 & 2$\times$323, 12$\times$348 & 2$\times$485, 2$\times$580, 2$\times$720 & & 5.1 \\
NGC 0288 & 00:52:45 & $-$26:34:43 & 6$\times$400 & 4$\times$350 & 4$\times$41                              &  6.3 \\
NGC 0362 & 01:03:14 & $-$70:50:53 & 6$\times$519 & 4$\times$350 & 4$\times$54                              &  6.3 \\
NGC 1261 & 03:12:16 & $-$55:12:58 & 4$\times$834, 4$\times$855, 2$\times$918 & 5$\times$413 & 5$\times$164 & 7.7  \\
NGC 1851 & 05:14:06 & $-$40:02:47 & 14$\times$1288 & 4$\times$453 & 2$\times$140                           &  5.7 \\
NGC 2298 & 06:48:59 & $-$36:00:19 & 4$\times$848, 4$\times$980 & 1$\times$327, 4$\times$350 & 5$\times$134 & 7.9  \\
NGC 2808 & 09:12:02 & $-$64:51:33 & 12$\times$985 & 6$\times$650 & 6$\times$97                             &  7.5 \\
NGC 3201 & 10:17:36 & $-$46:24:44 & 4$\times$754 & 4$\times$310 & 1$\times$60, 1$\times$68                 & 7.7  \\
NGC 4590 & 12:39:27 & $-$26:44:38 & 4$\times$696 & 4$\times$305 & 2$\times$60                              &  7.9 \\
NGC 4833 & 12:59:33 & $-$70:52:35 & 4$\times$895, 4$\times$936 & 4$\times$350 & 4$\times$135               & 7.7  \\
NGC 5024 & 13:12:55 &  18:10:05 & 3$\times$1729, 3$\times$1830 & 3$\times$433, 3$\times$460 & 3$\times$170, 3$\times$178 & 7.9 \\
NGC 5053 & 13:16:27 &  17:42:00 & 4$\times$829, 2$\times$854, 4$\times$879 & 5$\times$415 & 5$\times$164   & 8.0  \\
NGC 5272 & 13:42:11 &  28:22:31 & 6$\times$415 & 4$\times$350 & 4$\times$42                                & 6.2  \\
NGC 5286 & 13:46:26 & $-$51:22:27 & 2$\times$668, 2$\times$797 & 4$\times$310 & 2$\times$65                & 7.9  \\
NGC 5466 & 14:05:27 &  28:32:04 & 4$\times$865, 4$\times$928 & 4$\times$350 & 4$\times$135                 & 7.9  \\
NGC 5897 & 15:17:24 & $-$21:00:36 & 4$\times$874, 4$\times$926 & 4$\times$350 & 4$\times$138               & 8.0  \\
NGC 5904 & 15:18:33 &  02:04:51 & 4$\times$689 & 4$\times$306 & 2$\times$58                                & 8.1  \\ 
NGC 5927 & 15:28:00 & $-$50:40:26 & 6$\times$668 & 1$\times$30, 5$\times$310, 2$\times$475 & 2$\times$66   & 7.6  \\
NGC 5986 & 15:46:02 & $-$37:47:11 & 4$\times$668, 2$\times$745 & 6$\times$300 & 3$\times$60                & 8.3  \\
NGC 6093 & 16:17:02 & $-$22:58:30 & 10$\times$855 & 5$\times$657 & 5$\times$85                             & 6.2  \\
NGC 6101 & 16:25:48 & $-$72:12:07 & 4$\times$851, 4$\times$889, 2$\times$940 & 5$\times$415 & 5$\times$165 & 7.8  \\
NGC 6121 & 16:23:41 & $-$26:30:50 & 10$\times$380, 2$\times$735, 2$\times$808 & 4$\times$300 & 2$\times$66               & 6.9  \\
NGC 6144 & 16:27:13 & $-$26:01:24 & 4$\times$748 & 4$\times$304 & 2$\times$61                              & 8.0  \\
NGC 6171 & 16:32:31 & $-$13:03:13 & 4$\times$895, 4$\times$926 & 4$\times$350 & 4$\times$136               & 8.1  \\
NGC 6205 & 16:41:41 &  36:27:40 & 6$\times$427 & 4$\times$350 & 4$\times$46                                & 6.1  \\
NGC 6218 & 16:47:14 & $-$01:56:54 & 2$\times$714, 2$\times$790 & 4$\times$306 & 1$\times$58, 1$\times$66   & 7.9  \\
NGC 6254 & 16:57:09 & $-$04:06:01 & 2$\times$713, 2$\times$790 & 4$\times$306 & 2$\times$60                & 7.8  \\
NGC 6304 & 17:14:32 & $-$29:27:43 & 2$\times$693, 2$\times$800 & 4$\times$304 & 2$\times$60                & 7.8  \\
NGC 6341 & 17:17:07 &  43:07:58 & 2$\times$707, 2$\times$819 & 4$\times$304, 2$\times$425 & 2$\times$57    & 7.2  \\
NGC 6352 & 17:25:29 & $-$48:25:19 & 2$\times$706, 2$\times$800 & 4$\times$311, 5$\times$400, 1$\times$410 & 1$\times$58, 1$\times$72 & 7.0  \\
NGC 6362 & 17:31:54 & $-$67:02:52 & 2$\times$720, 2$\times$829 & 4$\times$323, 1$\times$368, 5$\times$450 & 2$\times$67     & 6.6  \\
NGC 6366 & 17:27:44 & $-$05:04:47 & 2$\times$713, 2$\times$795 & 4$\times$305 & 2$\times$62                              & 8.3  \\
NGC 6388 & 17:36:17 & $-$44:44:08 & 4$\times$888, 2$\times$961, 2$\times$999 & 4$\times$350 & 4$\times$133 & 8.2  \\
NGC 6397 & 17:40:35 & $-$53:39:57 & 2$\times$709, 2$\times$752 & 4$\times$310, 6$\times$620 & 2$\times$66  & 6.4  \\
NGC 6441 & 17:50:13 & $-$37:03:05 & 4$\times$887, 4$\times$928 & 4$\times$350 & 4$\times$123               & 8.0  \\
NGC 6496 & 17:59:03 & $-$44:15:57 & 2$\times$707, 2$\times$800 & 4$\times$303 & 1$\times$61, 1$\times$73   & 7.8   \\
NGC 6535 & 18:03:50 &  00:17:48 & 2$\times$713, 2$\times$793 & 1$\times$253, 4$\times$305, 5$\times$400 & 2$\times$62  & 6.8 \\
NGC 6541 & 18:08:02 & $-$43:42:53 & 2$\times$708, 2$\times$758 & 4$\times$300 & 2$\times$65                & 8.0  \\
NGC 6584 & 18:18:37 & $-$52:12:56 & 2$\times$709, 2$\times$795 & 4$\times$312 & 2$\times$62                & 7.6  \\
NGC 6624 & 18:23:40 & $-$30:21:39 & 2$\times$707, 2$\times$800 & 4$\times$295 & 1$\times$62, 1$\times$81   & 7.8  \\
NGC 6637 & 18:31:23 & $-$32:20:53 & 4$\times$887, 2$\times$923, 2$\times$933 & 4$\times$350 & 2$\times$120, 2$\times$130            & 7.9  \\
NGC 6652 & 18:35:45 & $-$32:59:26 & 2$\times$690, 2$\times$775, 2$\times$800 & 6$\times$305 & 1$\times$60, 1$\times$69, 1$\times$86 & 7.5  \\
NGC 6656 & 18:36:24 & $-$23:54:12 & 12$\times$812 & 4$\times$475 & 2$\times$141                            & 5.9  \\
NGC 6681 & 18:43:12 & $-$32:17:31 & 2$\times$706, 2$\times$800 & 4$\times$294 & 1$\times$66, 1$\times$83   & 7.7  \\
NGC 6715 & 18:55:03 & $-$30:28:47 & 3$\times$1751, 3$\times$1916 & 3$\times$433, 3$\times$475 & 3$\times$170, 3$\times$190          & 7.7  \\
NGC 6717 & 18:55:06 & $-$22:42:05 & 6$\times$686 & 6$\times$311 & 3$\times$60                              & 8.2  \\
NGC 6723 & 18:59:33 & $-$36:37:56 & 4$\times$693, 2$\times$734 & 6$\times$313 & 3$\times$60                & 8.0  \\
NGC 6752 & 19:10:54 & $-$59:59:11 & 12$\times$369 & 1$\times$30, 2$\times$500 &                            & 4.7  \\
NGC 6779 & 19:16:35 &  30:11:00 & 2$\times$706, 2$\times$800 & 4$\times$295 & 1$\times$64, 1$\times$81     & 7.8  \\
NGC 6791 & 19:20:52 &  37:46:18 & 4$\times$700 & 4$\times$297 &
2$\times$65                                & $  $\ \ 9.2$^\dagger$   \\
NGC 6809 & 19:39:59 & $-$30:57:53 & 2$\times$746 & 2$\times$294 & 1$\times$66                              & 7.9  \\
NGC 6838 & 19:53:46 &  18:46:45 & 2$\times$750, 2$\times$792 & 4$\times$303 & 2$\times$65                  & 7.7  \\
NGC 6934 & 20:34:11 &  07:24:16 & 2$\times$792 & 2$\times$304 & 1$\times$70                                & 7.5  \\
NGC 6981 & 20:53:27 & $-$12:32:14 & 2$\times$693, 2$\times$710 & 4$\times$304 & 2$\times$64                & 7.7  \\ 
NGC 7078 & 21:29:58 &  12:10:00 & 3$\times$615, 3$\times$700 & 6$\times$350 & 6$\times$65                  & 5.5  \\ 
NGC 7089 & 21:33:27 &  00:49:23 & 2$\times$676, 2$\times$735, 2$\times$785 & 6$\times$313 & 2$\times$62, 1$\times$70  & 7.4 \\
NGC 7099 & 21:40:22 & $-$23:10:47 & 2$\times$725 & 2$\times$303 & 1$\times$65                              & 8.1  \\

\tableline
\end{tabular}
\tablecomments{In addition to the GO-13297 exposures, this table
  includes the exposures from GO-12311 and GO-12605, the pilot 
  studies in  MPs using WFC3/UVIS.  Similar observations in
    the archive from
 programs GO-11633, GO-11729, GO-12008, GO-12746, and GO-12971 have
 also been included. 
 Exposure times are in seconds.\\
 $\Delta$Epoch corresponds to the time elapsed between first- and 
 second-epoch observations, namely the time baseline between the ACS GCS
  observations and the new UV WFC3/UVIS images.\\
 (“$^\dagger$) First epoch from GO-10265 (PI:\ T.\ Brown).
  }
\end{center}
\end{table*}

%
%

\vspace{0.7cm}

\section{Detecting and Measuring in a Common Frame}
\subsection{An automated reduction pipeline}
As was the case with the ACS GCS, the number of images used in this project 
and the desire for a homogeneous catalog necessitate the use of an automated 
reduction procedure.  This pipeline has to be effective enough to find stars 
(while avoiding artifacts) and at the same time produce results with a quality 
comparable to the original ACS GCS catalogs.  Below we describe the 
steps taken in order to accomplish this.

\subsection{Preliminary reduction steps}
All the WFC3/UVIS observations described in Table \ref{table:tab1} have 
been processed using the same procedures in order to provide homogeneous 
photometry.  WFC3/UVIS \texttt{\_flt} images have been corrected for the 
effects of 
imperfect charge-transfer efficiency (CTE)
using the parallelized 
version of the STScI publicly-available FORTRAN code\footnote{
     \url{http://www.stsci.edu/hst/wfc3/ins\_performance/CTE} } 
 (see Anderson \& Bedin 2010 for a detailed description of the method).
Stellar positions and fluxes for individual exposures 
were obtained 
using 
the program \texttt{hst1pass}, which is based on the publicly available
routine \texttt{img2xym\_WFC0.9x10} (Anderson \& King 2006) and will be made 
public along with a future paper (Anderson, in prep).  The code goes through 
each exposure, pixel-by-pixel, and identifies as a potential source of 
interest every pixel that:  (1) has at least 25 electrons of flux over sky 
in its brightest 2$\times$2 pixels and (2) is brighter than any other pixel 
within a radius of 3 pixels.  It uses an appropriate library point-spread 
function (PSF) to measure a least-squares position for the source 
(under the assumption that it is a star), 
a flux (using the central 5$\times$5 pixels and an aperture 
%
correction based on the star's location within its central pixel), and
also a quality-of-fit metric (Anderson et al. 2008).
Finally, the routine corrects the measured position for geometric distortion
using the solution in Bellini, Anderson, \& Bedin (2011), then uses 
the WCS header information to transform positions into a common frame 
(in this case, the ACS GCS frame for the cluster).

\begin{figure*}
\centering
\includegraphics[width=12cm]{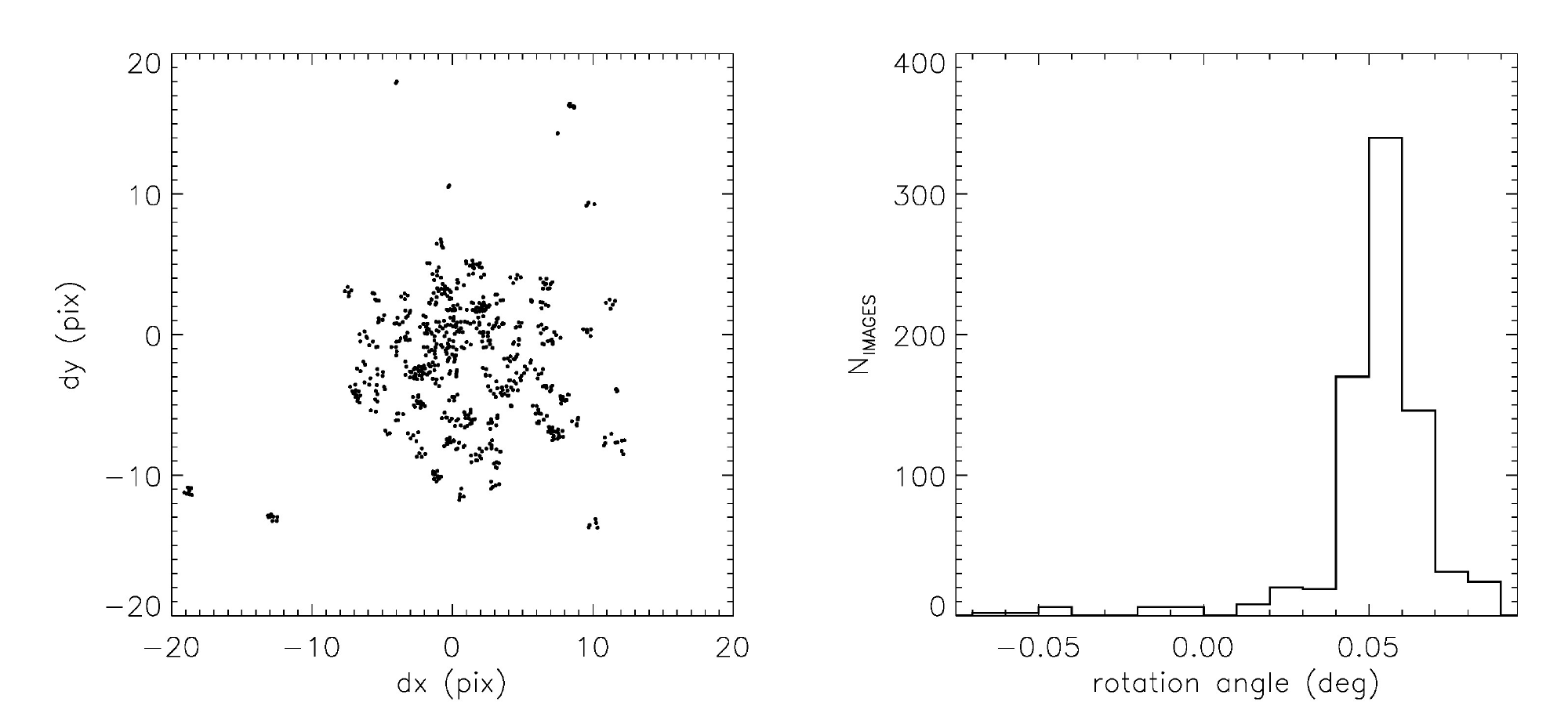}
\caption{
   \emph{Left},  distribution of offsets between our catalogs and the ACS GCS  
                 images derived from the 6-parameter linear transformations. 
                 Each point represents an exposure.  The four-point
                 clumpings 
%
                 reflect the fact that the dithered exposures in each visit tend 
                 to have the same guide-star related boresight offset.
   \emph{Right}, histogram showing the distribution of relative rotation 
                 angle between our catalogs and the ACS GCS.  The prevailing 
                 offset of 0.05 degree between the two frames indicates 
                 that our WFC3/UVIS distortion solution was not properly 
                 aligned with the V2 axis of the telescope.  This will be
                 improved before the release of the \texttt{hst1pass} software.
   \label{fig:hist}
   }
\end{figure*}

\begin{figure*}
\centering
\includegraphics[width=11cm]{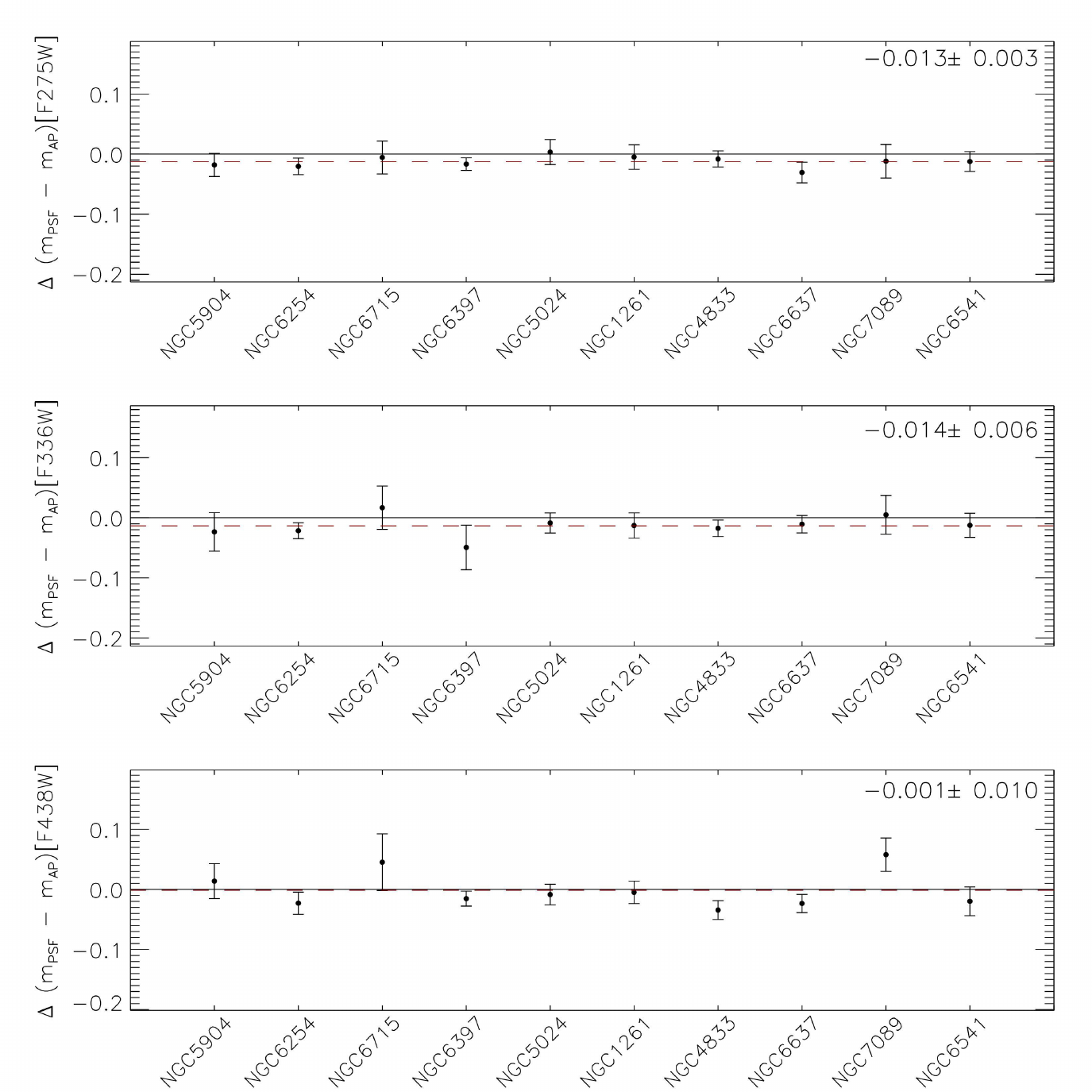}
\caption{Zeropoint difference between the PSF photometry of the
  \texttt{*\_flc} images and the aperture photometry of the \texttt{*\_drz} images
  using an aperture of 4.5 pixels (0\farcs18) in 10 GCs. These zeropoints include the
  correction for the normalization of our $uvm$ magnitudes to 1000 seconds. 
  \label{fig:zeropoint}
  }
\end{figure*}

\subsection{One-pass photometry}
A star must be detectable and measurable reliably on an individual 
exposure to be included in the present catalog.  The one-pass nature of 
this preliminary reduction does not allow us to go as faint as we will 
eventually go in the final catalog.  In the final catalog, we will find stars 
simultaneously on multiple exposures in multiple filters then measure them 
with consistent positions simultaneously in all available exposures, thus
allowing deeper photometry.  Here, we will restrict ourselves to the stars
that have at least S/N $\simeq$ 10 in individual exposures. 
%
In addition, since we are not attempting a full reduction of GO-13297 at
this time, we will simply cross-identify our starlists with the existing
ACS GCS catalog.  

%
The finding routine described above produces for each
image a list of sources, and for each source it determines an $x$ and $y$ 
position in the raw detector coordinates, a pixel-area-corrected raw 
instrumental magnitude $m$, an instrumental magnitude $m_{1000}$ that has been 
zeropointed to correspond to a 1000-s exposure, and a quality-of-fit 
metric $q$.  
The calibration of the instrumental $m_{1000}$ magnitudes into
VEGAMAG happens later in our procedure, as we will see in \S 3.5.
The routine also uses a distortion solution and the WCS 
information in the image header to transform the observed coordinates 
into the ACS GCS frame for that cluster to facilitate cross-identification.  
We call this reference frame the ($u$,$v$) frame.

\subsection{Cross-Identification with the ACS GCS}

The processing of the one-pass lists was performed by another routine,
named \texttt{uvm2collate}, which will also be described in Anderson
(in prep).  In brief, this routine cross-correlates the starlist from
each exposure with the Sarajedini catalog, while at the same time 
generating a representative stacked image in the reference frame for 
each filter.  

On account of typical errors of $\sim$0.5 arcsecond in the guide-star
positions that affect the absolute astrometry of all {\it HST} frames, we cannot
assume that the ($u$,$v$) positions constructed based on the WCS header of
each image to be in the ACS GCS frame will 
automatically
be perfectly aligned with the 
actual stars in the ACS GCS catalog.  
Thus, for each exposure, we take the 
($u$,$v$) list and find the horizontal and vertical offsets that provide 
the most matches with the relatively bright stars in the ACS GCS catalog 
%
using a matching radius of 1.75 pixels.  This generous radius makes our
initial matching robust against small errors in the absolute rotation 
of the {\it HST} frame as well.

We then take this initial cross-identification and use a full 6-parameter
linear transformation to relate our new positions to the master frame.
We compute residuals and iteratively reject the most discordant star until
there are no stars with residuals greater than 0.25 pixel.  We then use
these final transformations to compute improved ($u$,$v$) positions for
each cross-identified star.  Figure \ref{fig:hist} shows the distribution 
of offsets and orientation adjustments for the 780 images that were 
transformed in this process.  The offset represents the typical pointing
error 
resulting from
imperfect guide-star positions.  The prevailing orientation 
difference of 0.05 degree reflects the fact that the rotation we adopted 
between the distortion-solution frame of 
Bellini, Anderson, \& Bedin (2011) 
and the V2 axis was slightly 
off.  This did not prevent all the stars from being properly matched, 
and it will be fixed by the time the \texttt{hst1pass} software is released.

Once the final transformation from each exposure into the reference frame
has been established, we take the original loosely cross-identified list 
and determine improved reference-frame positions for each star.  The routine
then takes all the cross-identified stars and computes an average position 
(and an RMS about that position) for each star in the GO-13297 data set 
and an average instrumental magnitude (and an RMS about that magnitude) 
in each band, iteratively rejecting observations that are discordant at 
the 3-sigma level.  The position that we compute here is not optimized 
for astrometry; to do that we would have to take more care in performing 
the transformations and would have to weight the observations according to 
signal-to-noise.  
Even with these caveats, the position should be accurate to 
 at least the 0.1 pixel-level  
(5 mas, implying a proper motion error of about 0.6
mas/year,
 with a time baseline of 7-9 years).  
This should 
enable cluster-field rejection, even though it is not adequate for a study 
of the internal motions.  A high-precision internal-motion catalog will 
be constructed using the algorithm in Bellini et al. (2014), using all
available observations.


\subsection{Photometric Calibration}
Thus far in the procedure, our photometry of the GO-13297 images has been
kept in the raw, instrumental system in order to give us a direct idea of
the intrinsic signal-to-noise of each measurement.  We also retain the 
1000-s exposure-time-zeropointed photometry to facilitate eventual absolute 
calibration.  While the PSFs we used to measure the one-pass fluxes were 
normalized to correspond to the total within 10 pixels ($0\farcs4$),  
there is distortion in the raw frames and thus there can be small offsets 
between our photometry and that produced on the \texttt{drizzle} products,
which have been carefully calibrated to VEGAMAG zeropoints.

Therefore, in order to do the absolute calibration, we first had to reduce 
the DRZ images with aperture photometry using the standard 10-pixel-radius 
aperture, add the official zeropoints\footnote{
 \url{http://www.stsci.edu/hst/wfc3/phot\_zp\_lbn}},
and the {\tt \_drz-aperture} corrections\footnote{\url{http://www.stsci.edu/hst/wfc3/documents/handbooks/currentIHB/c06_uvis07.html}},
 and cross-identify these stars 
with the ACS GCS master list in order to arrive at a calibrated
magnitude (i.e., in VEGAMAG)
for each star detected in each UVIS band.  The next step was to determine 
a zeropoint for each filter.

The difference between our 1000-s-normalized magnitudes and the calibrated
photometry should be a simple zeropoint that is the same for all clusters,
since the process that goes from \texttt{\_flt} to \texttt{\_drz} is the
same linear process for all images.  
Therefore, for ten representative
clusters, we determined the offset between our photometry and the VEGAMAG
photometry.  Figure~\ref{fig:zeropoint} shows the zeropoint determined
 for the three different filters for ten different clusters.
The average zeropoint is noted at the top.  We have used the average 
of the zeropoint correction for these 10 clusters 
for the entire sample of 57 GCs.
  
\subsection{Improvements to come}
It bears repeating that the catalog presented here is preliminary.  
It is based on a one-pass reduction with static library PSFs and 
it contains only those stars that were already found in Sarajedini's 
GC Treasury project.  The CMDs shown in Piotto et al. (2015) and 
papers based on them are different from those that can be constructed 
from these catalogs
in that they are based on a more careful treatment 
of each cluster (including differential reddening corrections, 
custom PSF modeling for each exposure, UVIS-based star lists, etc).  
All of these improvements will eventually be available for the final 
public catalog, but until all clusters can be reduced in the final
optimal way, we decided to make this preliminary, uniformly reduced
catalog available now.  Section 5 will detail the anticipated improvements.


\begin{figure*}
\centering
\includegraphics[width=\textwidth]{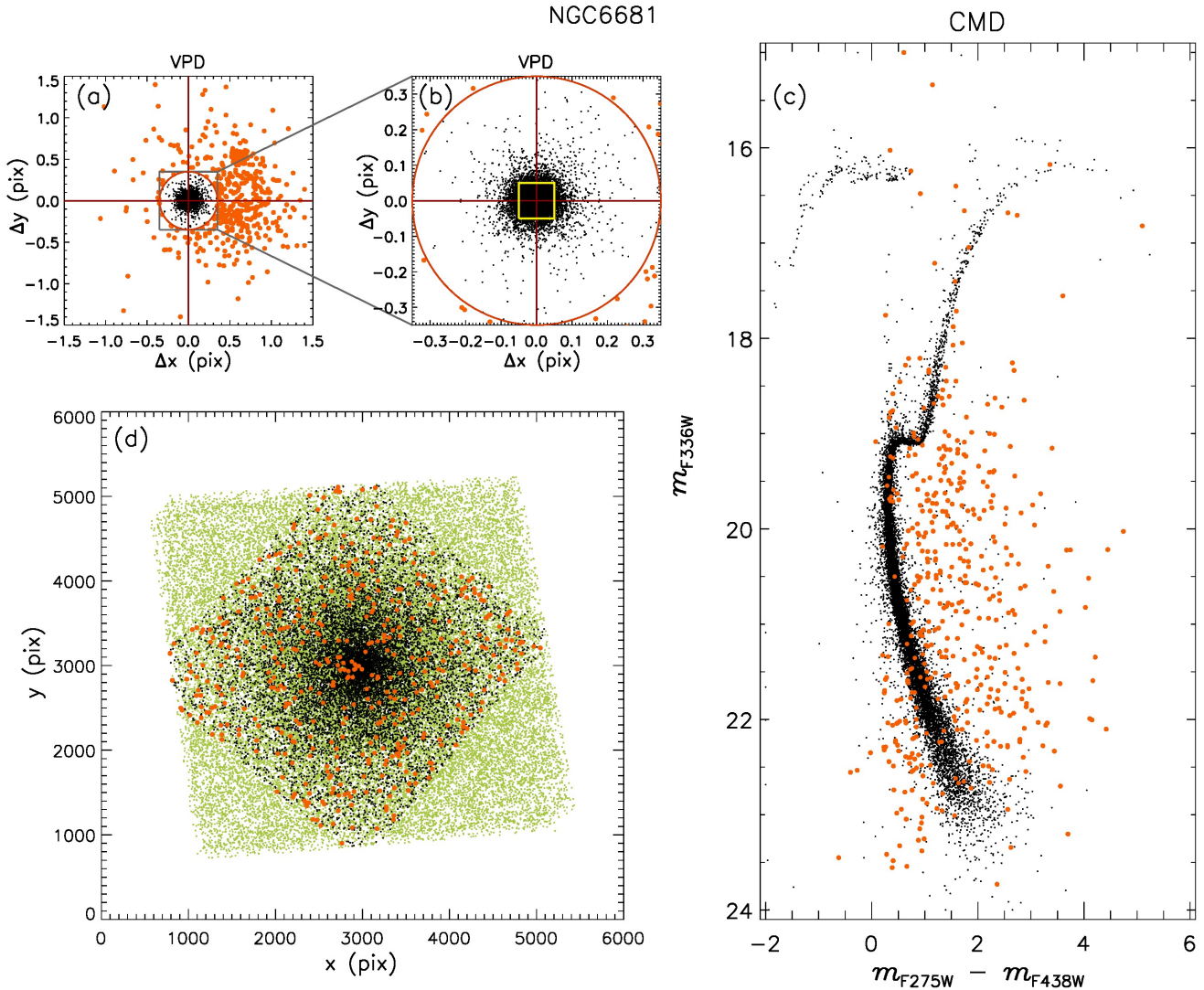}
\caption{Illustration of the data that our catalog provides for the GC NGC~6681.
         Panel (a) shows the vector-point diagram (VPD), the displacement (in ACS/WFC pixels), between the 2006 ACS GCS 
        and this catalog (2013-2014) --- a $\sim$7-8 yr time-baseline.
         We have marked stars with a displacement greater than 
         0.35 pixel (orange circle) as field stars (orange points) in 
         all plots. Panel (b) is a zoomed version of (a), where the 
         central yellow square represents the reliability of our current 
         crude 2013 positions ($0\farcs005$). Panel (c) shows the UV CMD 
         for the complete field, while Panel (d) shows the positions of 
         stars detected in the ACS GCS (light green points) overplotted with 
         those that have also been detected in our WFC3/UVIS UV catalogs 
         (black points)
 \label{fig:panel41}
}
\end{figure*}


\section{Legacy Products}

The collation procedure described above operates on the list of stars in
the original ACS GCS catalog for each cluster.  It determines an average 
position for each star in the reference frame based on the observations 
from all the filters along with an average magnitude and rms about that 
magnitude for each star in each filter.  It also reports for each filter 
the number of exposures in which a star could be found ({\texttt p}), 
the number of times it was found ({\texttt f}), and the number of times 
it was found consistently and not rejected ({\texttt g}).  

\subsection{Catalog Products}
These quantities all come from the new GO-13297 data and are reported 
in the first 17 columns of our catalog for each cluster.  
Table~ \ref{tab:table2} lists all the columns in our catalogs.  For 
convenience, Columns 18 through 24 provide basic information 
from the ACS GCS to allow easy construction of five-band photometry and 
cluster-field separation proper motions.
 In columns 25 and 26 we list the displacements which we have rounded to 0.1 pixels.

Figure \ref{fig:panel41} shows an example of our catalog products for
cluster NGC~6681.  Panels (a) and (b) compare the positions of each 
star in the ACS GCS and the average position in our F275W, F336W 
and F438W catalog, which reflects a time-baseline of $\sim$ 7-8\ years. 
Panel (c) shows the color-magnitude diagram (CMD) in our UV bands, 
while panel (d) shows the physical distribution of the stars in the 
ACS GCS reference frame. 

 
In panel (b) we have drawn a square representing the expected 
astrometric accuracy of $0.1$ pixel centered on the origin.  Similarly, 
we have marked in panels (a) and (b) the stars outside a radius of $0.35$ pixel
as field stars (orange points).  The same field stars have been 
highlighted in panels (c) and (d).



\subsection{Image Products}

In addition to producing the catalog, our procedure also uses the inverse 
transformations and inverse distortion solutions to map each pixel of the 
individual \texttt{\_flc} exposures into the reference frame to generate 
a representative stack of each band in the ACS GCS frame.  These stacks 
represent the sigma-clipped mean of the nearest pixels from the contributing
images and provide a more visual representation of the GO-13297 data in the 
same frame as the F606W images of the ACS GCS reduction.  This should enable 
users to quickly assess whether particular stars that were missed in this 
preliminary reduction will be likely to be included when we do the more 
comprehensive multi-pass final reduction.  

\begin{figure*}
\centering
\includegraphics[width=\textwidth]{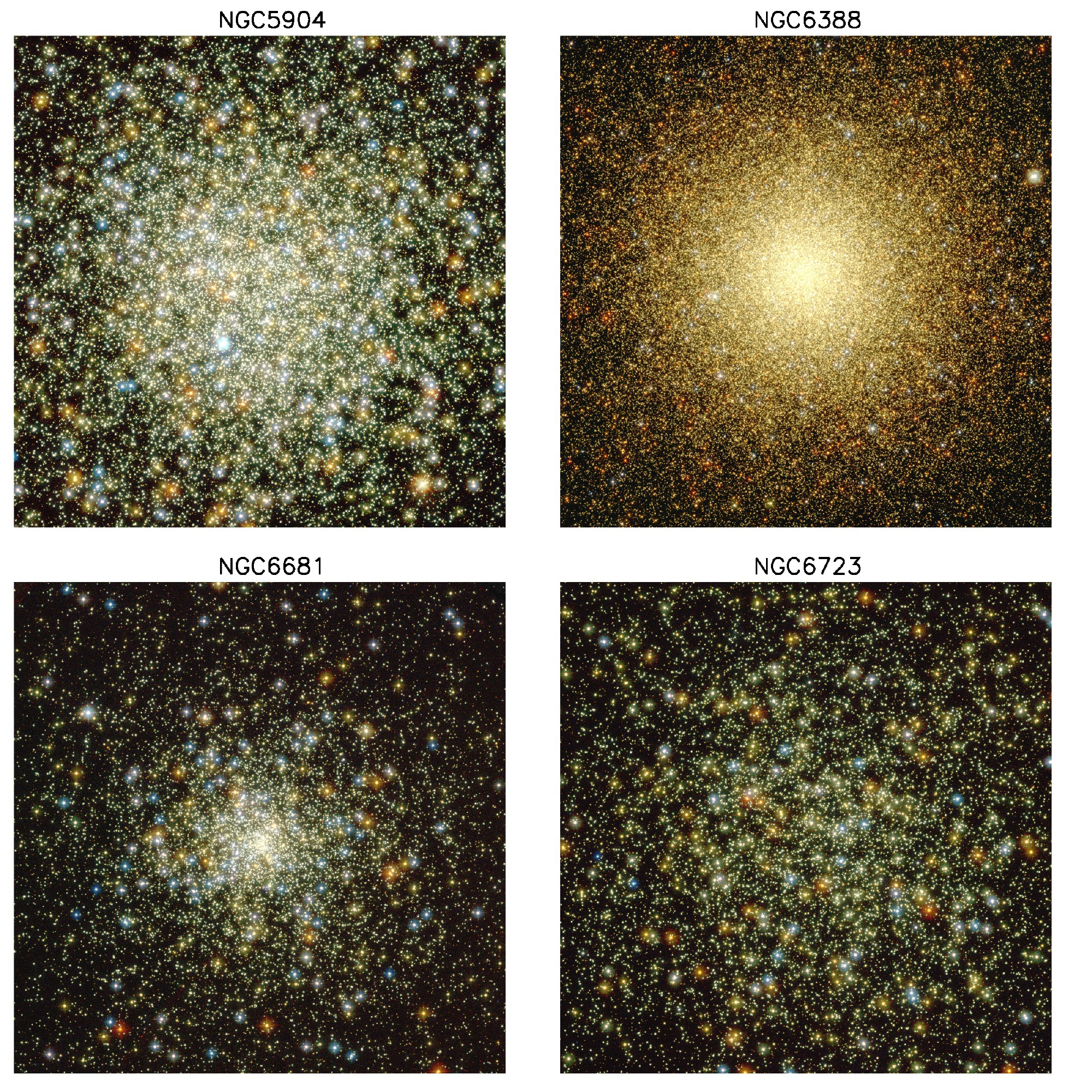}
\caption{Trichromatic stack images of 4 clusters in our survey 
         NGC5904, NGC6388, NGC6681 and NGC6723, each image covers 
         a field of $1\farcm8\times1\farcm8$.
\label{fig:rgb}
}
\end{figure*}

These stacks are normalized to correspond to a 1000-s exposure and have 
the same WCS header information as the GO-10775 images (which have been 
astrometrically corrected for {\it HST} guide-star errors to 2MASS).  We also 
place into the image header the pixel location of the cluster center as determined 
by Goldsbury et al.\ (2010).  Finally, we combined together the F275W, F336W, 
and F438W images into a single RGB TIFF image for each cluster.  This 
is not useful for science, but 
may be useful for presentations
and highlight the extreme populations.  Figure \ref{fig:rgb} 
shows the stacked 
color
images for four different clusters.

Eventually, these data will be uploaded to the Mikulski Archive for
Space Telescopes\footnote{\url{http://archive.stsci.edu}} (MAST), but until then they
can be accessed at
\url{http://www.astro.uda.cl/public_release/globularclusters41.html}
and at \url{http://groups.dfa.unipd.it/ESPG/treasury.php}.

\section{Summary}
This brief paper describes the preliminary catalog we have generated
for the 48 clusters of program GO-13297, the \emph{Hubble Space Telescope 
UV Legacy Survey of Galactic Globular Clusters}.  
We also 
include in the catalog data taken with previous programs that had the 
same filter set and scientific goals, for a total of 57 clusters.

We ran a one-pass photometry routine on each individual exposure and
collated the results against the catalog from GO-10775, the 2006-epoch 
ACS survey in F606W and F814W.  We provide to the community calibrated 
photometry for each star found in F275W, F336W, and F438W, along with 
empirical uncertainties and a 2013-epoch position, which can be used for 
cluster-field separation.  Color-color plots from these new catalogs 
should be particularly effective for identifying stars in different 
populations to enable 
spectroscopic follow-up; this is
the main reason we sought to make this preliminary catalog available
before the final reduction.  In addition, we produce a representative 
stacked image in each band along with a color-composite image.

The data provided here represent the preliminary data products for the
survey.  The photometry here is different from that in Piotto et al. 
(2015) in that no customization has been done for each cluster in terms
of reddening correction, PSF-modeling, etc.  The goal here was to provide
a uniform reduction to the community.

Over the next year, we will do a more comprehensive reduction of the
dataset.  Having the entire dataset in hand will enable us to construct
an improved model of the PSF based on the entire dataset, allowing
for variations due to focus.  Even with such an improved PSF model, we
will also construct a perturbation PSF for each exposure so that the
stars can be measured as accurately and as consistently as possible 
(Faint stars are best measured with smaller apertures, but without a good
PSF model, this can introduce systematic errors; to avoid this, here 
we had to use the same 5$\times$5-pixel aperture for all stars).  
We will also improve the distortion solutions and independently 
evaluate the astrometric and photometric errors that remain after 
the pixel-based CTE correction.

The final reduction will not be a one-pass procedure as was done here;
rather it will use a multi-pass procedure similar to that described
in Anderson et al.\ (2008), 
though 
optimized for the three-filter dataset
that we have here.  We will produce new stacks along with subtracted stacks
to help users evaluate whether there may be stars that were just below the
one-pass finding threshold.  We will also perform artificial-star tests 
to allow users to evaluate the completeness, photometric errors and 
possible recovery biases.  Finally, we will use the final dataset 
to construct a reddening correction specific to the local neighborhood 
of each star.  These improvements should be made over the course of the next 
year, with the catalogs released to the community soon thereafter.



\acknowledgments

MS, JA, AB, AC, and IRK acknowledge support from STScI grant GO-13297. 
M.S. acknowledges support from Becas Chile de Postdoctorado en el Extranjero 
project  74150088. GP, VG, SO, and DN acknowledge partial support  by the
Universit\`a degli Studi di Padova Progetto di Ateneo CPDA141214
``Towards understanding complex star formation in Galactic globular
clusters'' and by INAF under the program PRIN-INAF2014.
 A.A., S.C., and S.H. recognize partial support by the IAC 
(grant P301031) and the Ministry of Competitiveness and Innovation of Spain 
(grant AYA2010-16717). APM acknowledges support by the Australian 
Research Council through Discovery Early Career Researcher Award
 DE150101816.  
\afterpage{\clearpage}
\tiny
\begin{turnpage}
\begin{table*}
\setlength{\tabcolsep}{3pt}
\caption{Catalogue example: NGC6681 \label{tab:table2}}
\begin{tabular}{ccc ccc c c c c c c c c c c c c c c c c c c r r}
\tableline\tableline
 xuv  & yuv &  m275  &  m336  &  m438  & m275$\_$sig & m336$\_$sig& m438$\_$sig &  g2&  f2&  p2&   g3 & f3 & p3 &  g4 & f4 & p4& id$\_$ata & x$\_$ata  &y$\_$ata & m606 &m814 & RA$\_$ata& DEC$\_$ata & D$\_$x & D$\_$y \\   
\tableline
      0.0    &  0.0 &  0.000 &  0.000 &  0.000 &  0.000  & 0.000 &  0.000 &  0 &  0  & 0 &   0 &  0 &  0 &   0  & 0 &  0 &  00001 & 1103.67 &  776.87 & 20.534 & 19.676 & 280.83379 & $-$32.32285 & 0.0 & 0.0 \\
      0.0    &  0.0 &  0.000 &  0.000 &  0.000 &  0.000  & 0.000 &  0.000 &  0 &  0  & 0 &   0 &  0 &  0 &   0  & 0 &  0 &  00002 & 1090.91 &  984.99 & 19.944 & 19.344 & 280.83400 & $-$32.31995 & 0.0 & 0.0 \\
      0.0    &  0.0 &  0.000 &  0.000 &  0.000 &  0.000  & 0.000 &  0.000 &  0 &  0  & 0 &   0 &  0 &  0 &   0  & 0 &  0 &  00003 & 1140.71 &  809.17 & 21.593 & 20.817 & 280.83318 & $-$32.32240 & 0.0 & 0.0 \\
      0.0    &  0.0 &  0.000 &  0.000 &  0.000 &  0.000  & 0.000 &  0.000 &  0 &  0  & 0 &   0 &  0 &  0 &   0  & 0 &  0 &  00004 & 1167.88 &  842.12 & 20.588 & 19.954 & 280.83273 & $-$32.32194 & 0.0 & 0.0 \\
      0.0    &  0.0 &  0.000 &  0.000 &  0.000 &  0.000  & 0.000 &  0.000 &  0 &  0  & 0 &   0 &  0 &  0 &   0  & 0 &  0 &  00005 & 1178.94 &  850.65 & 20.542 & 19.821 & 280.83255 & $-$32.32182 & 0.0 & 0.0 \\
   2582.9 &  3653.3 & 22.703 & 21.575 & 21.787 &  0.132  & 0.045 &  0.083 &  3 &  3 &  4  &  4  & 4 &  4  &  2 &  2 &  2 &  32554 & 2582.96 & 3653.30 & 20.862 & 20.198 & 280.80947 & $-$32.28290 & 0.0 & 0.0 \\
   2584.3 &  3661.3 & 20.582 & 19.840 & 20.271 &  0.027  & 0.092 &  0.025 &  4 &  4 &  4  &  4  & 4 &  4  &  2 &  2 &  2 &  32555 & 2584.31 & 3661.32 & 19.538 & 18.988 & 280.80945 & $-$32.28279 & 0.0 & 0.0 \\
   2586.3 &  3657.9 & 21.652 & 19.917 & 19.844 &  0.048  & 0.033 &  0.003 &  4 &  4 &  4  &  4  & 4 &  4  &  2 &  2 &  2 &  32556 & 2586.52 & 3657.77 & 18.735 & 18.042 & 280.80941 & $-$32.28284 & $-$0.3 & 0.2 \\
   2582.0 &  3679.3 & 20.171 & 19.560 & 19.900 &  0.106  & 0.013 &  0.003 &  4 &  4 &  4  &  4  & 4 &  4  &  2 &  2 &  2 &  32557 & 2581.94 & 3679.37 & 19.158 & 18.619 & 280.80949 & $-$32.28254 & 0.0 & 0.0 \\
   2566.6 &  3700.3 &  0.000 &  0.000 &  0.000 &  0.000  & 0.000 &  0.000 &  0 &  0 &  4  &  0  & 0 &  4  &  0 &  0 &  2 &  32558 & 2566.55 & 3701.00 & 26.305 & 24.527 & 280.80974 & $-$32.28224 & 0.1 & $-$0.7 \\
   2564.8 &  3713.2 & 20.260 & 19.625 & 19.924 &  0.046 &  0.021 &  0.015 &  4 &  4 &  4 &   4 &  4 &  4 &   2 &  2 &  2 &  32559 & 2564.83 & 3713.17 & 19.194 & 18.655 & 280.80977 & $-$32.28207 & 0.0 & 0.0 \\
   ...   &  ...       &   ...      &  ...  &  ...  &  ... & ...  &  ... &  ...  & ...  & ... &  ...  &  ... & ... & ... & ...  & ...  &  ...  & ...  & ...  & ...  & ... & ... & ... & ... & ... \\
\tableline                  
\end{tabular}
\tablecomments{Each column description can be summarized as follows:\\
xuv,yuv are the coordinates transformed into the 0.05\arcsec/pix ACS GCS frame \\
m275 is the VEGAMAG calibrated magnitude in F275W band\\
m336 is the VEGAMAG calibrated magnitude in F336W band\\
m438 is the VEGAMAG calibrated magnitude in F438W band\\ 
m275$\_$sig is the photometric error of magnitude in F275W filter \\
m336$\_$sig is the photometric error of magnitude in F336W filter \\
m438$\_$sig is the photometric error of magnitude in F438W filter \\
gi, fi, pi are the number of times a star has: a good measurement (g), is found(f), and \\
could have been found (p). The subindex indicates the band of the measurement,\\ 
F275W ($i=2$), F336W ($i=3$), F438W ($i=4$)\\
id$\_$ata is the identification number of the star in the ACS GCS \\
x$\_$ata, y$\_$ata are the pixel coordinates of the star in ACS GCS, the coordinates \\
 span from 0 to 6000 pixels in each axis\\
m606, m814 are the VEGAMAG calibrated magnitudes in the ACS GCS in the F606W and F814W filters\\
RA$\_$ata, DEC$\_$ata are the Right Ascension and Declination in the original ACS GCS \\
D$\_$x, D$\_$y are the displacements in pixels, in the original ACS GCS axes and pixel size, these \\
are calculated using the ACS GCS and new WFC3 UV positions as first and second epoch respectively.\\  
The displacements  can be converted into mas\,yr$^{-1}$ proper motions using the formula \\
PM$_{i}$= (D$\_$i $\times$ 50.0)/$\Delta$Epoch, where the
$\Delta$Epoch can be found in Table 1 and $i$=x$,$y.\\
 The obtained proper motions can be expressed in units of km\,s$^{-1}$ using \\
  Vel[km\,s$^{-1}$]= 4.74 $\times$ dist[kpc] $\times$
  PM[mas\,yr$^{-1}$], with dist corresponding to the distance \\
of the cluster from the Sun. We must stress that, while the provided displacements\\
 D$\_$i have an accuracy sufficient  for cluster-field separation, they are not suited for\\
 internal cluster dynamics. \\
}
\end{table*}
\end{turnpage}

\afterpage{\clearpage}

\end{document}